\documentclass[twocolumn]{emulateapj}

\usepackage{epsfig}

\usepackage{bm}

\usepackage{graphicx}
\usepackage{subfigure}
\usepackage{url}

\usepackage{amssymb}
\usepackage{times}
\usepackage[usenames]{color}

\usepackage{xcolor}
\usepackage[colorlinks=true, urlcolor=blue]{hyperref}

\usepackage{natbib}
\usepackage{rotating}
\usepackage{color}
\def\cyan#1 {\textcolor{cyan}{#1}~}

\journalinfo{{\sc Submitted to The Astrophysical Journal Letters}}
\submitted{Submitted to ApJL *Fill In*}

\shorttitle{POST-OUTBURST RADIO OBSERVATIONS OF PSR J1119-6127}
\shortauthors{MAJID ET AL.}

\begin{document}

% Setup hyperlinks.
\hypersetup{
	colorlinks,
	linkcolor={blue!50!black},
	citecolor={blue!50!black},
	urlcolor={blue!50!black}
}

\title{Post-Outburst Radio Observations of the High Magnetic Field Pulsar PSR J1119-6127}

\newcommand{\Majid}{Walid~A.~Majid}
\newcommand{\Pearlman}{Aaron~B.~Pearlman}
\newcommand{\Dobreva}{Tatyana~Dobreva}
\newcommand{\Kocz}{Jonathon~Kocz}
\newcommand{\Lippuner}{Jonas~Lippuner}
\newcommand{\Prince}{Thomas~A.~Prince}
\newcommand{\Shinji}{Shinji~Horiuchi}

\author{\Majid\altaffilmark{1,2}, \Pearlman\altaffilmark{2,1,5,6}, \Dobreva\altaffilmark{1}, \Shinji\altaffilmark{4}, \Kocz\altaffilmark{1}, \Lippuner\altaffilmark{3}, \Prince\altaffilmark{2,1}}

\newcommand{\CaltechPhysics}{Division of Physics, Mathematics, and Astronomy, California Institute of Technology, Pasadena, CA 91125, USA}
\newcommand{\JPL}{Jet Propulsion Laboratory, California Institute of Technology, Pasadena, CA 91109, USA}
\newcommand{\TAPIR}{TAPIR, Walter Burke Institute for Theoretical Physics, MC 350-17, California Institute of Technology, Pasadena, CA 91125, USA}
\newcommand{\CSIRO}{CSIRO Astronomy and Space Science, Canberra Deep Space Communications Complex, PO Box 1035, Tuggeranong, ACT 2901, Australia}
\newcommand{\NDSEG}{NDSEG Research Fellow}
\newcommand{\NSF}{NSF Graduate Research Fellow}

\address{
$^{1}$ \JPL \\
$^{2}$ \CaltechPhysics \\
$^{3}$ \TAPIR \\
$^{4}$ \CSIRO \\
$^{5}$ \NDSEG \\
$^{6}$ \NSF
}

\def\psrA{PSR~J1119-6127}
\def\psrB{PSR~J1846-0258}

\begin{abstract}

We have carried out high frequency radio observations of the high magnetic field pulsar PSR~J1119-6127 following its recent X-ray outburst. While initial observations showed no evidence of significant radio emission, subsequent observations detected pulsed emission across a large frequency band. In this letter, we report on the initial disappearance of the pulsed emission and its prompt reactivation and dramatic evolution over several months of observation. The periodic pulse profile at S-band (2.3\,GHz) after reactivation exhibits a multi-component emission structure, while the simultaneous X-band (8.4\,GHz) profile shows a single emission peak. Single pulses were also detected at S-band near the main emission peaks. We present measurements of the spectral index across a wide frequency bandwidth, which captures the underlying changes in the radio emission profile of the neutron star. The high frequency radio detection, unusual emission profile, and observed variability suggest similarities with magnetars, which may independently link the high energy outbursts to magnetar-like behavior.

\end{abstract}


%%%%%%%%%%%%%%%%%%%%%%%%%%%%%%%%%%%%%%%%%%%%%%%%%%%%%%%%%%%%%%

\section{Introduction}

PSR~J1119-6127 is a young radio pulsar with a spin period of $P = \text{0.410}$\,s and a period derivative of $\dot{P} = \text{4.0}\,\times\,\text{10}^{\text{--12}}$, which is one of the highest reported spin-down rates for a radio pulsar. The pulsar has a characteristic age of 1.6\,kyr and an inferred surface dipole magnetic field of $B = \text{4.1}\,\times\,\text{10}^{\text{13}}$\,G, one of the largest among known radio pulsars. PSR~J1119-6127 was initially discovered in the Parkes multibeam pulsar survey~\citep{Camillo2000} and is likely associated with the Galactic supernova remnant G292.2-0.5~\citep{Crawford2001} at a distance of 8.4 kpc~\citep{Caswell2004}. This pulsar has been detected in X-rays~\citep{Gonzalez2003} and gamma-rays~\citep{Parent2011}, and it is also known to glitch~\citep{Weltevrede2011}. Unusual pulse profile changes, short radio bursts, and irregular timing recoveries~\citep{Weltevrede2011, Antonopoulou2015} have been observed following a glitching event.

On 2016 July 27 13:02:08 UT~\citep{Younes2016} and 2016 July 28 01:27:51 UT~\citep{Kennea2016}, short magnetar-like bursts from PSR~J1119-6127 were detected from the \textit{Fermi} Gamma-Ray Burst Monitor (GBM) and the \textit{Swift} Burst Alert Telescope (BAT), respectively. Soon after the announcement of the BAT outburst, the \textit{Swift} X-ray Telescope (XRT) detected a bright X-ray source at the position of this pulsar with pulsed emission at a pulse period of 0.4098627(3)\,s, consistent with the known rotational period of the pulsar~\citep{Antonopoulou2016}. Prior to the outburst, X-ray pulsed emission was detected only in the soft band ($<$\,2.5\,keV). XRT measurements following the outburst showed strong pulsations with a pulsed fraction of 60\% across the XRT energy band spanning 2.5--10\,keV. A glitch was reported by~\citet{Archibald2016b} using \textit{Swift} XRT, \textit{NuSTAR}, and \textit{Fermi} Large Area Telescope (LAT) data. \citet{Archibald2016a} observed spectral hardening of PSR~J1119-6127 following the high energy outburst, which is suggestive of magnetar-like emission as in the case of the rotation-powered pulsar PSR~J1846-0258~\citep{Gavriil2008}. \citet{Gogus2016} uncovered a total of 12 hard X-ray bursts during 2016 July 26-28 using \textit{Fermi} GBM and \textit{Swift} XRT observations and carried out spectral and temporal analyses of the emission.

Adding to this unusual behavior of PSR~J1119-6127, immediate radio follow-up of the pulsar at 1465\,MHz (L-band) using the Parkes radio telescope on two consecutive days (2016 July 29 starting at 04:59:00 UT and 2016 July 30 starting at 01:08:18 UT) failed to detect radio pulsations, placing an upper limit of 90\,$\mu$Jy~\citep{Burgay2016a} on the flux density of this pulsar. Given that the typical flux density at the same frequency is about 1\,mJy, with a few percent fluctuations at most, the disappearance of pulsed emission at this frequency implied a reduction of the radio flux by more than a factor of 10 after the high energy outburst. Reactivated radio pulsations were detected on 2016 August 09 03:40:12 UT via continued monitoring of the pulsar at the Parkes telescope~\citep{Burgay2016b}.

In this letter, we report our results using high-frequency Target of Opportunity (ToO) observations of PSR~J1119-6127 at the 70-m Deep Space Network (DSN) antenna in Canberra (DSS-43), Australia, carried out both before and after the return of pulsed emission from this pulsar.

%%%%%%%%%%%%%%%%%%%%%%%%%%%%%%%%%%%%%%%%%%%%%%%%%%%%%%%%%%%%%%

\section{Radio Observations}
\label{Section:Observations}

Following the reported X-ray outburst, we observed PSR~J1119-6127 on four separate epochs with DSS-43. This antenna is equipped with cryogenically cooled dual polarization receivers centered at 2.3 GHz (S-band) and 8.4 GHz (X-band), arranged so that both bands can be used simultaneously with their beams concentric on the sky. The four output signals are then sent to a newly commissioned ultra-wideband pulsar machine, which was specially developed to meet the requirements of searching for short period pulsars at high frequencies with wide bandwidths.  The pulsar machine is a digital filterbank system that is capable of processing 16 independent input bands, each up to 1 GHz wide. Spectra for each band have 1024 channels and can be produced with time sampling as short as 32\,$\mu$s. These are then individually recorded to disk for further processing.
  
In the observations reported here, we used the pulsar machine with four input bands and simultaneously recorded both S/X-bands in dual circular polarization mode. On-off measurements of a standard calibrator, Hydra A (3C218), were carried at the start of each observation, which yielded an estimated system temperature of 25/40 K (20\% error) at S/X-band.
The antenna gain was $\sim$1\,K/Jy. The data used in this study spanned 96\,MHz and 480\,MHz at S-band and X-band, respectively. The data were recorded with a frequency channel spacing of 1\,MHz and time sampling of 512\,$\mu$s at 16 bits per sample. In Table~\ref{Table:RadioObservations}, we list all four observing epochs with their start times and durations. While epochs 1 and 2 were carried out prior to the reported reactivation of radio pulsed emission~\citep{Burgay2016b}, the latter two epochs were taken after this reactivation.

As reported in~\cite{Majid2016}, observations during the initial two epochs failed to detect any evidence of either periodic emission or single pulses. Using the calibration data and system configuration, we obtained upper limits of 0.14/0.06\,mJy for pulsed emission at S/X-band. The ATNF pulsar catalog\footnote{See http://www.atnf.csiro.au/research/pulsar/psrcat.}~\citep{Manchester2005} lists flux densities of 0.80/0.44\,mJy at 1.4 and 3\,GHz, respectively, yielding a relatively flat spectral index of -0.8 over this frequency range. Using this spectral index, the expected flux densities at S/X-band before the X-ray outburst were 0.6/0.2\,mJy, respectively. We inferred that the pulsed emission at these higher frequencies during the first two epochs was suppressed by a factor of~3 or more at S-band and X-band following the magnetar-like outburst event. We also searched for single pulses with widths up to 130\,ms and detected none with a signal-to-noise ratio (SNR) above~6.0. We placed an upper limit of 45/20\,mJy on the flux densities of bright single pulses at S/X-band.

Periodicity and single pulse searches were also performed using radio observations during epochs~3 and~4 after the pulsed radio emission was reactivated. We first searched the data for evidence of narrow-band and wide-band radio frequency interference (RFI) using mild filtering criteria, which required only 5\% of the data to be discarded. The contaminated portion of the data was masked and removed from our analysis using the \texttt{rfifind} tool from the \texttt{PRESTO}\footnote{See http://www.cv.nrao.edu/$\sim$sransom/presto.} pulsar search package~\citep{Ransom2001}. We also corrected for the bandpass slope across the frequency band and removed the baseline using a high-pass filter with a time constant of~1\,s. The data were then dedispered using a dispersion measure (DM) of 707.4\,pc\,cm$^{\mathrm{-3}}$~\citep{He2013}. A problem with the calibration system in epoch~3 prevented us from using data from the RCP (right circular polarization) channel at S-band. For consistency and to avoid introducing any systematic errors, we only used a single polarization channel for all epochs in the analysis presented here. In Section~\ref{Section:Results}, we present the results of periodicity searches and the discovery of single pulses at S-band during these latter epochs.

%%%%%%%%%%%%%%%%%%%%%%%%%%%%%%%%%%%%%%%%%%%%%%%%%%%%%%%%%%%%%%

\begin{deluxetable}{ccccc}
	\footnotesize
	\tablecaption{Radio Observations of PSR J1119-6127 with the DSN}
	\tablewidth{0pt}
	\tablehead{
		\colhead{Epoch} &
		\colhead{Date$^{\mathrm{a}}$} &
		\colhead{Time$^{\mathrm{a}}$} &
		\colhead{Date$^{\mathrm{b}}$} &
		\colhead{Duration} \\
		\colhead{} &
		\colhead{} &
		\colhead{(hh:mm:ss)} &
		\colhead{(MJD)} &
		\colhead{(hr)}
	}
	\startdata
	1 & 2016 Jul 31 & 23:41:12 & 57600.98694 & 0.6 \\
	2 & 2016 Aug 01 & 23:37:12 & 57601.98417 & 3.7 \\
	3 & 2016 Aug 19 & 18:27:34 & 57619.76914 & 5.3 \\
	4 & 2016 Sep 01 & 17:31:34 & 57632.73025 & 3.2
	\enddata
	\tablecomments{\\
		$^{\mathrm{a}}$ Start time of the observation (UTC). \\
		$^{\mathrm{b}}$ Start time of the observation.}
	\label{Table:RadioObservations}
\end{deluxetable}

%%%%%%%%%%%%%%%%%%%%%%%%%%%%%%%%%%%%%%%%%%%%%%%%%%%%%%%%%%%%%%

\section{Results}
\label{Section:Results}

\subsection{Pulse Profiles \& Spectral Index}
\label{Section:Pulse_Profiles}
 
\begin{figure*}[t]
	\centering
	\begin{tabular}{ccc}
		
		\subfigure
		{
			\includegraphics[trim=0cm 0cm 0cm 0cm, clip=false, scale=0.4, angle=0]{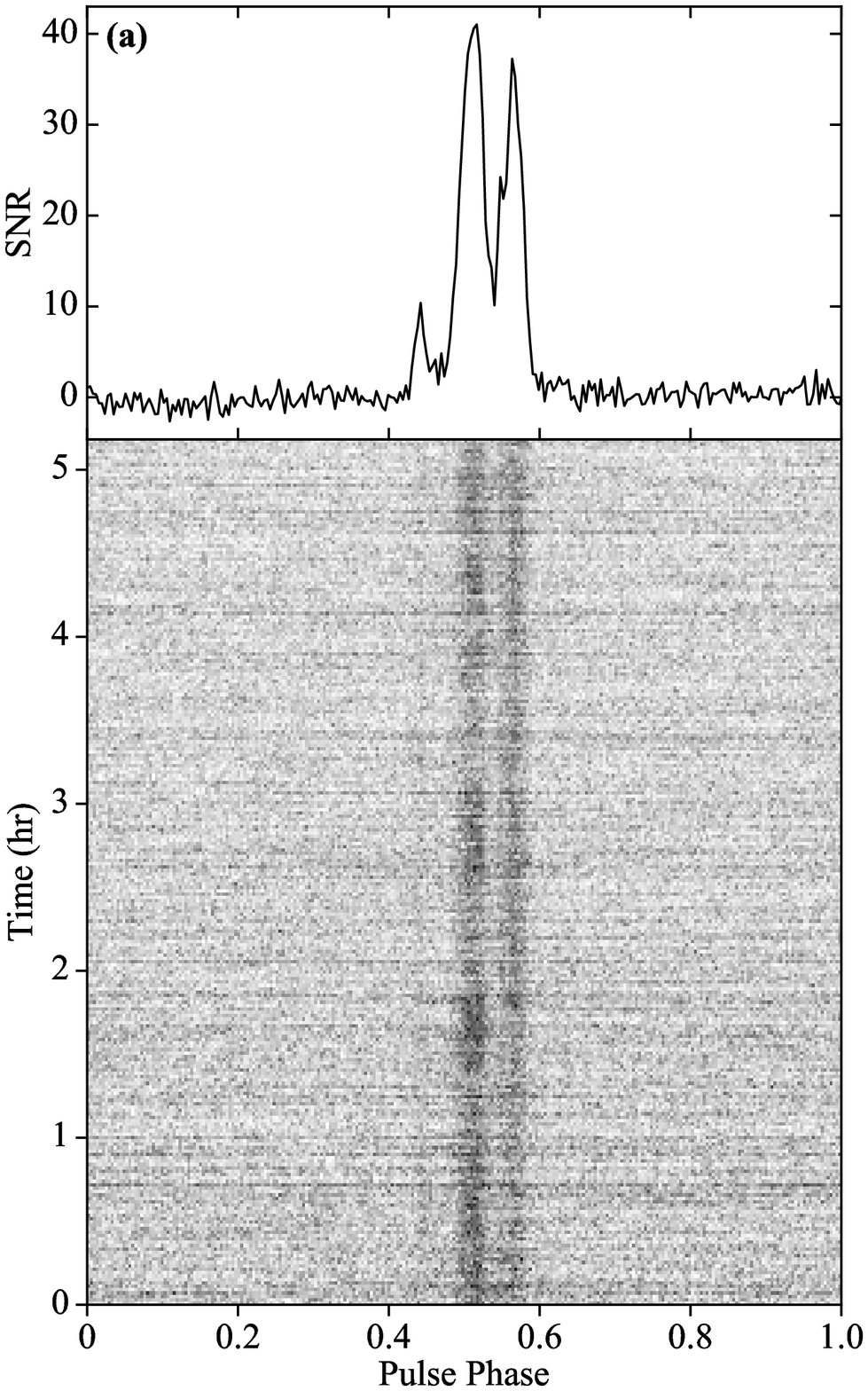}
			\label{Figure:Figure1a}
		}
		
		& &
		
		\subfigure
		{
			\includegraphics[trim=0cm 0cm 0cm 0cm, clip=false, scale=0.4, angle=0]{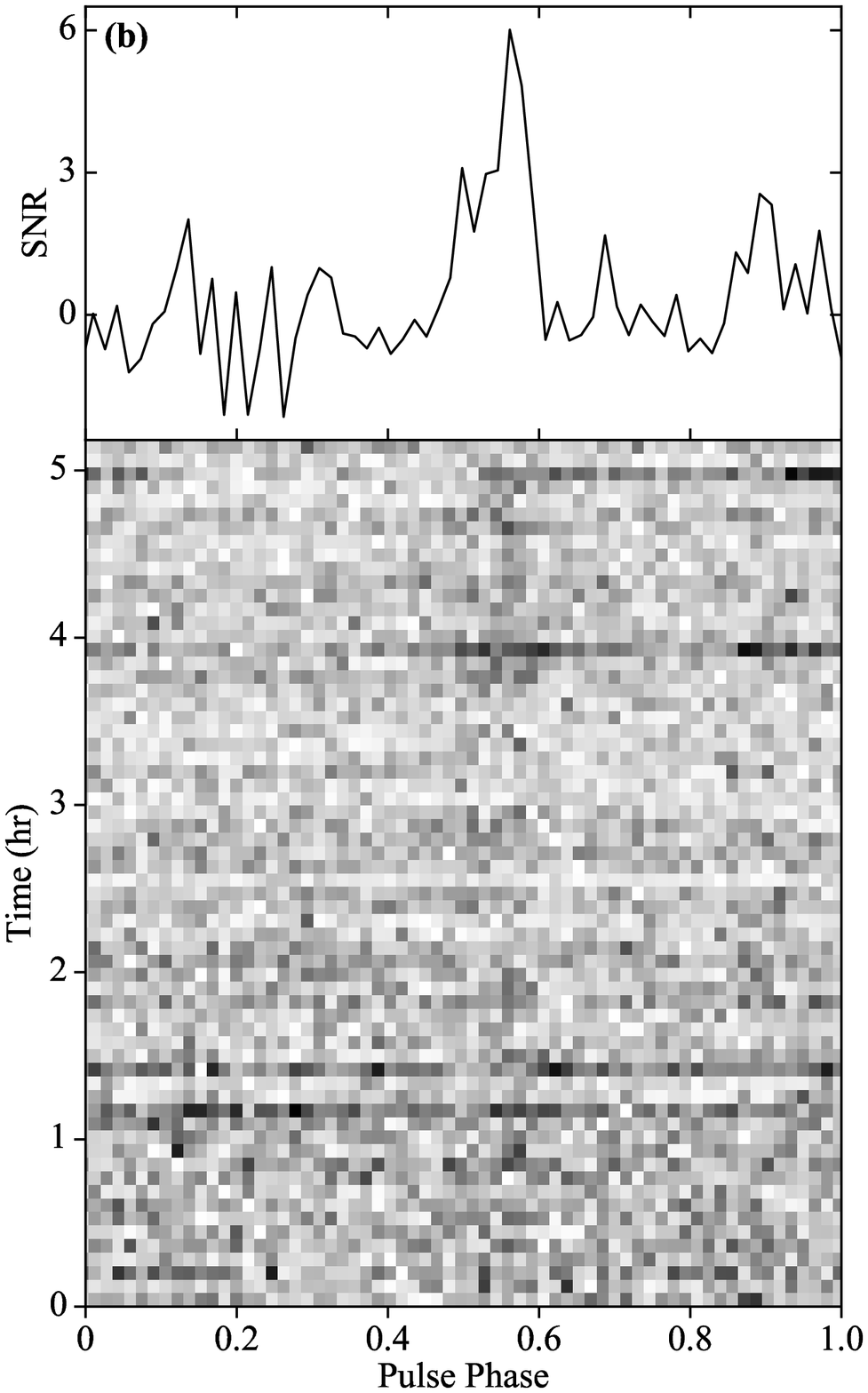}
			\label{Figure:Figure1b}
		}
		
		\\
		
		\subfigure
		{
			\includegraphics[trim=0cm 0cm 0cm 0cm, clip=false, scale=0.4, angle=0]{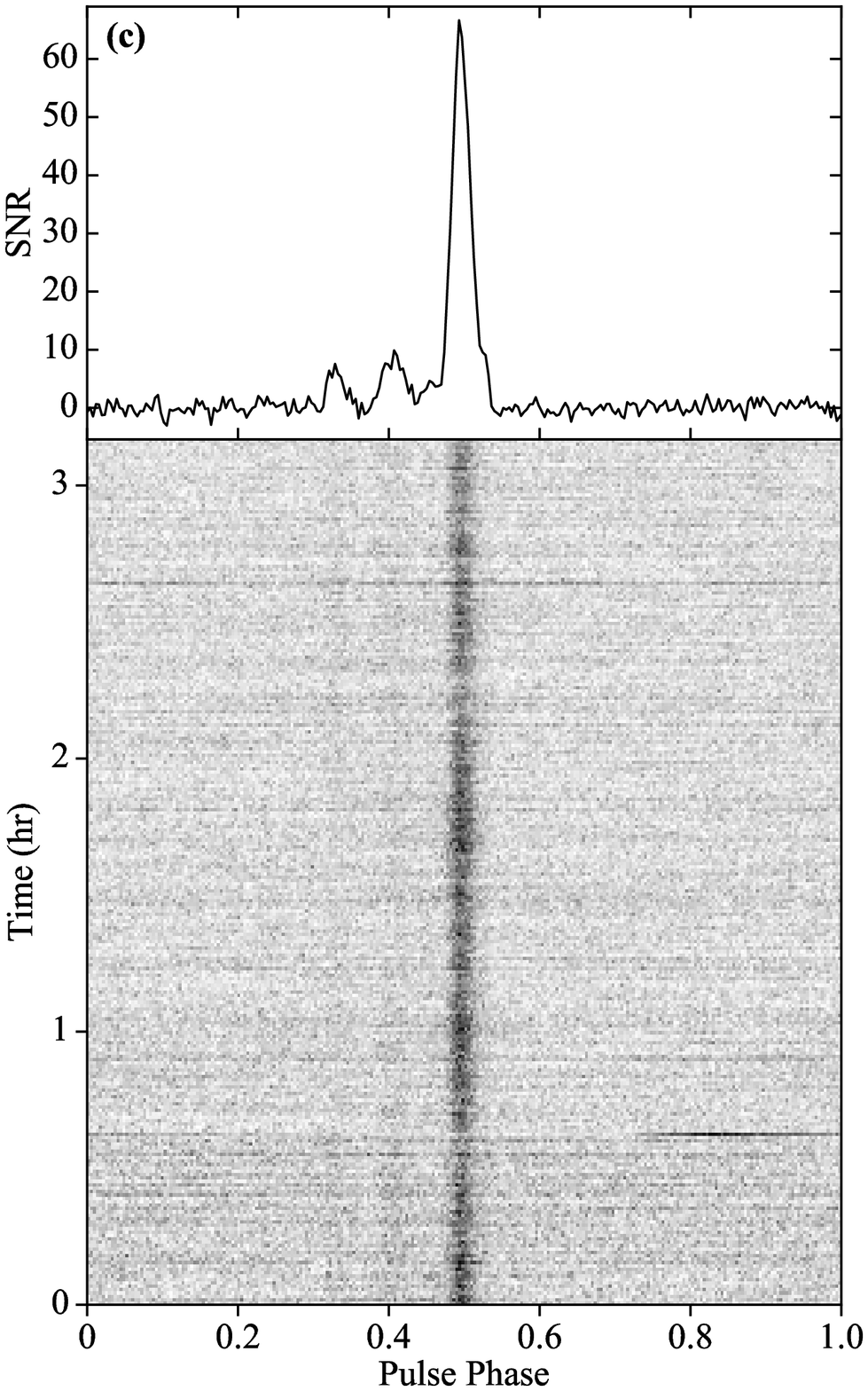}
			\label{Figure:Figure1c}
		}
		
		& &
		
		\subfigure
		{
			\includegraphics[trim=0cm 0cm 0cm 0cm, clip=false, scale=0.4, angle=0]{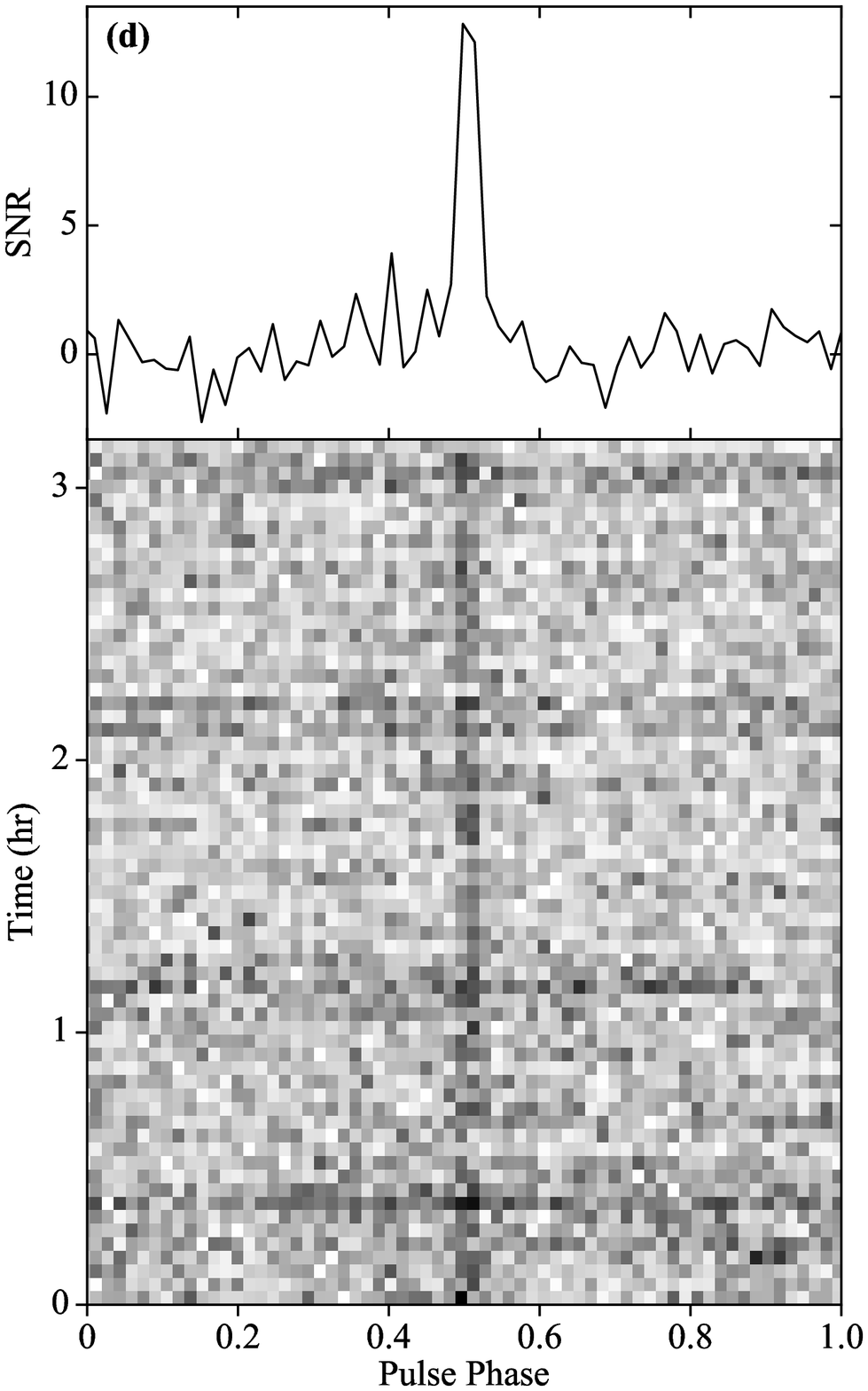}
			\label{Figure:Figure1d}
		}
		
	\end{tabular}
	
	\caption{Pulse profiles of PSR~J1119-6127 during epoch~3 (top row) and epoch~4 (bottom row) at S-band (left column) and X-band (right column). The top panels show the integrated pulse profiles in units of SNR and the grey-scale bottom panels show the strength of the pulsations as a function of phase and time, where darker bins correspond to stronger pulsed emission. The number of phase bins is 256/64 in the S/X-band profiles.}
	\label{Figure:Figure1}
\end{figure*}

Periodicity searches were performed using the \texttt{PRESTO} pulsar search package~\citep{Ransom2001} near the spin period of the pulsar with an updated ephemeris provided by~\cite{Archibald2016a}. We observed bright pulsed emission at S-band during epochs~3 and~4 after the reported reactivation of radio emission at L-band~\citep{Burgay2016b}. We find best barycentric pulse periods of 409.87281(3)\,ms and 409.88631(3)\,ms from S-band observations during epochs 3 and 4, respectively. 
Pulse profiles were obtained after barycenteric and DM corrections were applied at each epoch.
In Figure~\ref{Figure:Figure1}, we show the detection of PSR~J1119-6127 at S-band and X-band during epochs~3 and~4 after the radio emission was reactivated. The profiles during epoch~3 and epoch~4 were not aligned since the barycentric pulse period changed between the two epochs.

The pulse profile of PSR~J1119-6127 had been observed to be single peaked at 1.37\,GHz~\citep{Camillo2000} before the recent high energy outburst. However, the S-band pulse profile during epoch~3 in Figure~\ref{Figure:Figure1a} shows a triple-peaked structure, with two prominent peaks following a pre-cursor emission region. We continued to observe significant changes in the profile shape on the time scale of days. By epoch~4, the S-band pulse profile in Figure~\ref{Figure:Figure1c} had evolved into a strong single-peak, with two of the peaks from epoch~3 diminishing in strength. We marginally detected the pulsar at X-band over a 5 hour observing period in epoch~3, while significant X-band pulsed emission was observed during epoch~4, as shown in Figure~\ref{Figure:Figure1b} and Figure~\ref{Figure:Figure1d}, respectively.

We estimate the observed S-band flux densities from the LCP (left circularly polarized) channel taken on epochs~3 and~4 to be $S_{\mathrm{2.3}} = \mathrm{1.8(4)}$\,mJy and $S_{\mathrm{2.3}} = \mathrm{2.0(4)}$\,mJy, respectively. The S-band flux density has increased by at least a factor of 10 over the two week period of inactivity in pulsed emission. From observations taken on epoch~3, we estimate a mean X-band flux density of $S_{\mathrm{8.4}} = \mathrm{0.10(2)}$\,mJy, yielding a spectral index of $\alpha = \mathrm{-2.2(2)}$, where $S_{\nu} \propto \nu^{\alpha}$. Data taken on epoch~4 shows an increase in X-band flux density, $S_{\mathrm{8.4}} = \mathrm{0.18(4)}$\,mJy, yielding a slightly flatter spectral index of $\alpha = \mathrm{-1.9(2)}$. The ATNF pulsar catalog$^{\text{1}}$ ~\citep{Manchester2005} reports a spectral index of -0.8 based on flux density measurements at 1.4 and 3\,GHz prior to the outburst. To our knowledge, there are no reported flux density measurements of PSR~J1119-6127 at or near 8\,GHz prior to the recent outburst. If we assume that the reported spectral index is valid up to $\sim$8\,GHz frequency, then our measurements indicate a steepening of the spectral index. Our measured spectral index values are in the range expected among the population of radio pulsars, which has a mean index of -1.4~\citep{Bates2013} with standard deviation of order unity.

%%%%%%%%%%%%%%%%%%%%%%%%%%%%%%%%%%%%%%%%%%%%%%%%%%%%%%%%%%%%%%

\begin{deluxetable}{ccccc}
	\footnotesize
	\tablecaption{Flux Density and Spectral Index Measurements}
	%\tablewidth{0pt}
	\tablehead{
		\colhead{Epoch} &
		\colhead{$S_{\mathrm{2.3}}$$^{\mathrm{a}}$} &
		\colhead{$S_{\mathrm{8.4}}$$^{\mathrm{b}}$} &
		\colhead{$\alpha$$^{\mathrm{c}}$} \\
		\colhead{} &
		\colhead{(mJy)} &
		\colhead{(mJy)} &
		\colhead{}
	}
	\startdata
	1 & $<$\,0.4(1) & $<$\,0.24(5) & \nodata \\
	2 & $<$\,0.14(3) & $<$\,0.06(1) & \nodata \\
	3 & 1.8(4) & 0.10(2) & -2.2(2) \\
	4 & 2.0(4) & 0.18(4) & -1.9(2) \\
	\enddata
	\tablecomments{\\
		$^{\mathrm{a}}$ Mean flux density at 2.3\,GHz. \\
		$^{\mathrm{b}}$ Mean flux density at 8.4\,GHz. \\
		$^{\mathrm{c}}$ Spectral index from 2.3\,GHz to 8.4\,GHz.}
	\label{Table:FluxDensity}
\end{deluxetable}

%%%%%%%%%%%%%%%%%%%%%%%%%%%%%%%%%%%%%%%%%%%%%%%%%%%%%%%%%%%%%%

% Figure 1 - Pulse Profiles (S-band and X-band).

%%%%%%%%%%%%%%%%%%%%%%%%%%%%%%%%%%%%%%%%%%%%%%%%%%%%%%%%%%%%%%

\subsection{Single Pulses}
\label{Section:Single_Pulse_Search}
 
We present an analysis of the distribution of S-band single pulses during epochs~3 and~4 after the radio emission was reactivated.
The raw data for each observation epoch in Table~\ref{Table:RadioObservations} were barycentered and dedispersed using a DM of 707.4\,pc\,cm$^{\mathrm{-3}}$ after applying a mask to remove RFI. The dedispered data was then smoothed, normalized, and searched using a matched filtering algorithm, where the full resolution data was convolved with boxcar kernels of varying widths. 
We searched for individual, bright pulses~\citep{Weltevrede2011} with widths up to 100\,ms and SNRs above~4.0 in the time domain using the \texttt{PRESTO} pulsar search software package~\citep{Ransom2001}. \texttt{PRESTO} calculates the SNR of a candidate pulse using:  

\begin{equation}
\text{SNR} = \frac{\sum_{i}(S_i - M)}{\sigma \sqrt{w}},
\end{equation}

\noindent where the sum is over successive bins $S_i$ in the boxcar function, $M$ is local mean, $\sigma$ is the root-mean-square noise after normalization, and $w$ is the boxcar width in number of bins. The data were already normalized such that $M$ $\approx$ 0 and $\sigma \approx 1$. This definition of SNR has the advantage that it gives
approximately the same result regardless of the downsampling factor used for the time series~\citep{Deneva2016}.

Single pulses that were found using a DM of 707.4\,pc\,cm$^{\mathrm{-3}}$, which were coincident with events obtained without correcting for dedispersion, were removed from the candidate list. We also excluded candidates that were falsely identified as a result of our procedure for masking RFI. We chose to restrict our candidate list to pulses with widths less than 16\,ms since many of the single pulses with larger widths were determined to be RFI events from their dynamic spectra, which showed the strength of the pulses in frequency and time. No bright single pulses were detected in epochs~1 and~2 before the reappearance of the pulsed radio emission or at X-band during any of the epochs in Table~\ref{Table:RadioObservations}.

\begin{figure*}[t]
	\centering
	\begin{tabular}{ccc}
		
		\subfigure
		{
			\includegraphics[trim=0cm 0cm 0cm 0cm, clip=false, scale=0.4, angle=0]{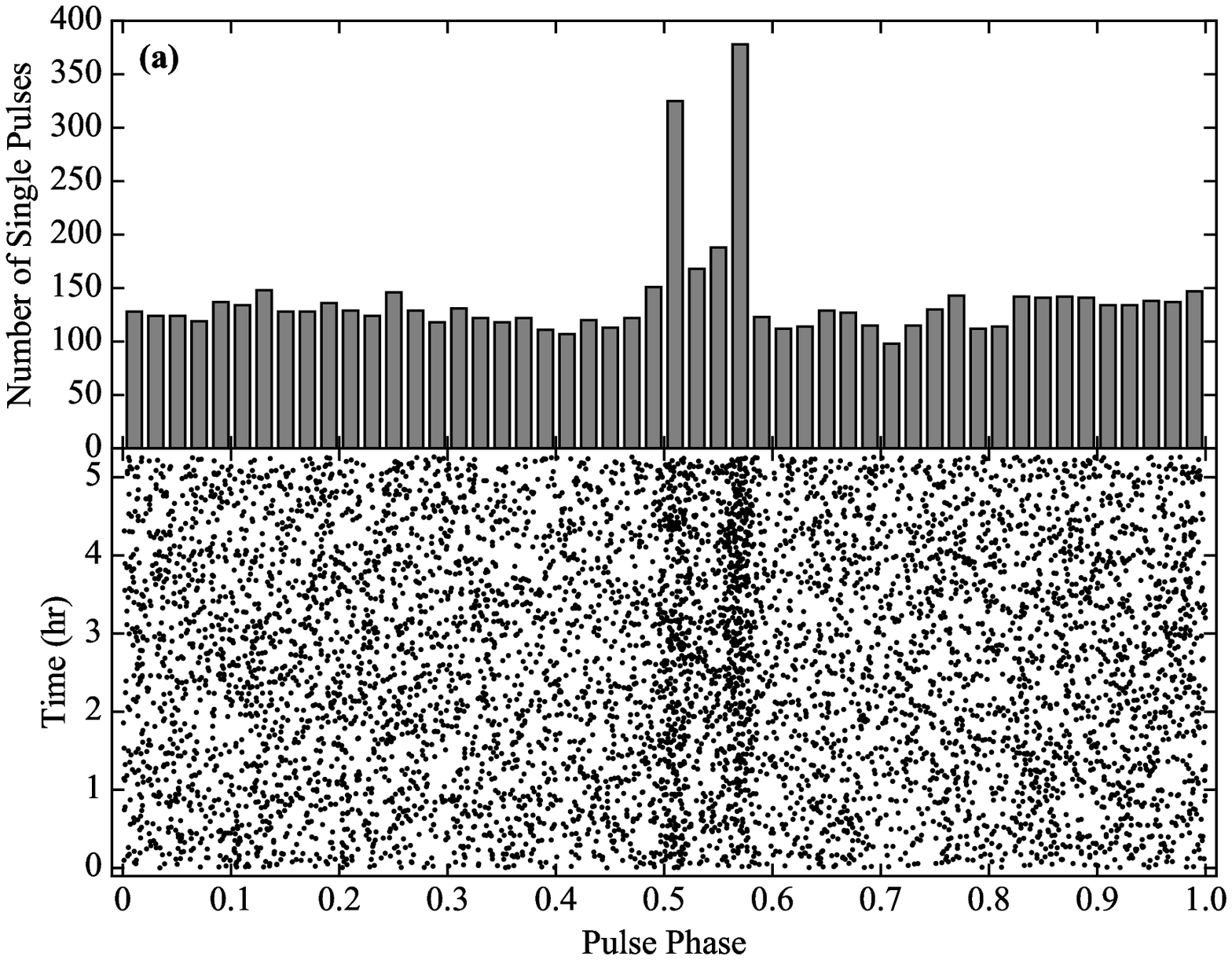}
			\label{Figure:Figure2a}
		}
		
		& &
		
		\subfigure
		{
			\includegraphics[trim=0cm 0cm 0cm 0cm, clip=false, scale=0.4, angle=0]{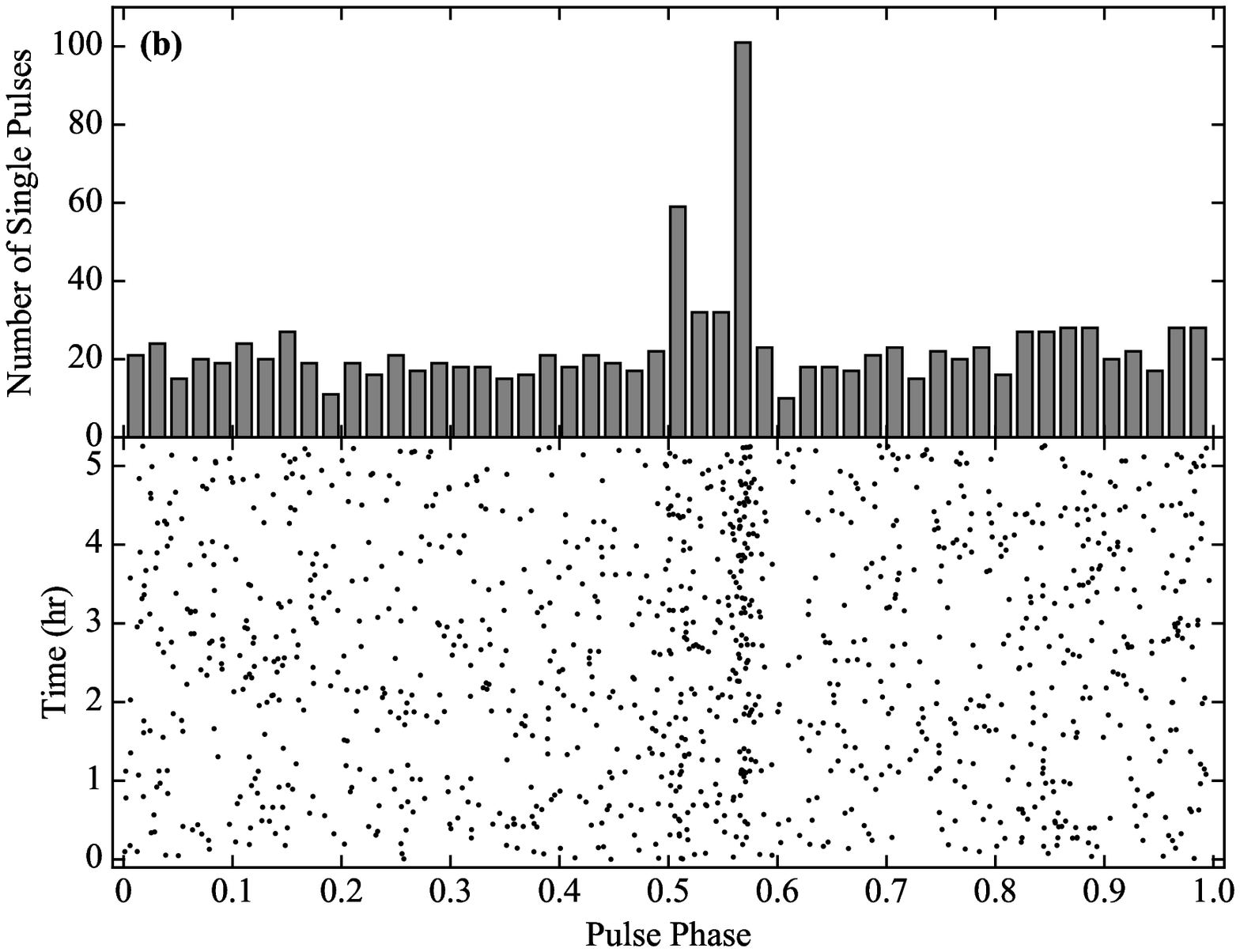}
			\label{Figure:Figure2b}
		}
		
		\\
		
		\subfigure
		{
			\includegraphics[trim=0cm 0cm 0cm 0cm, clip=false, scale=0.4, angle=0]{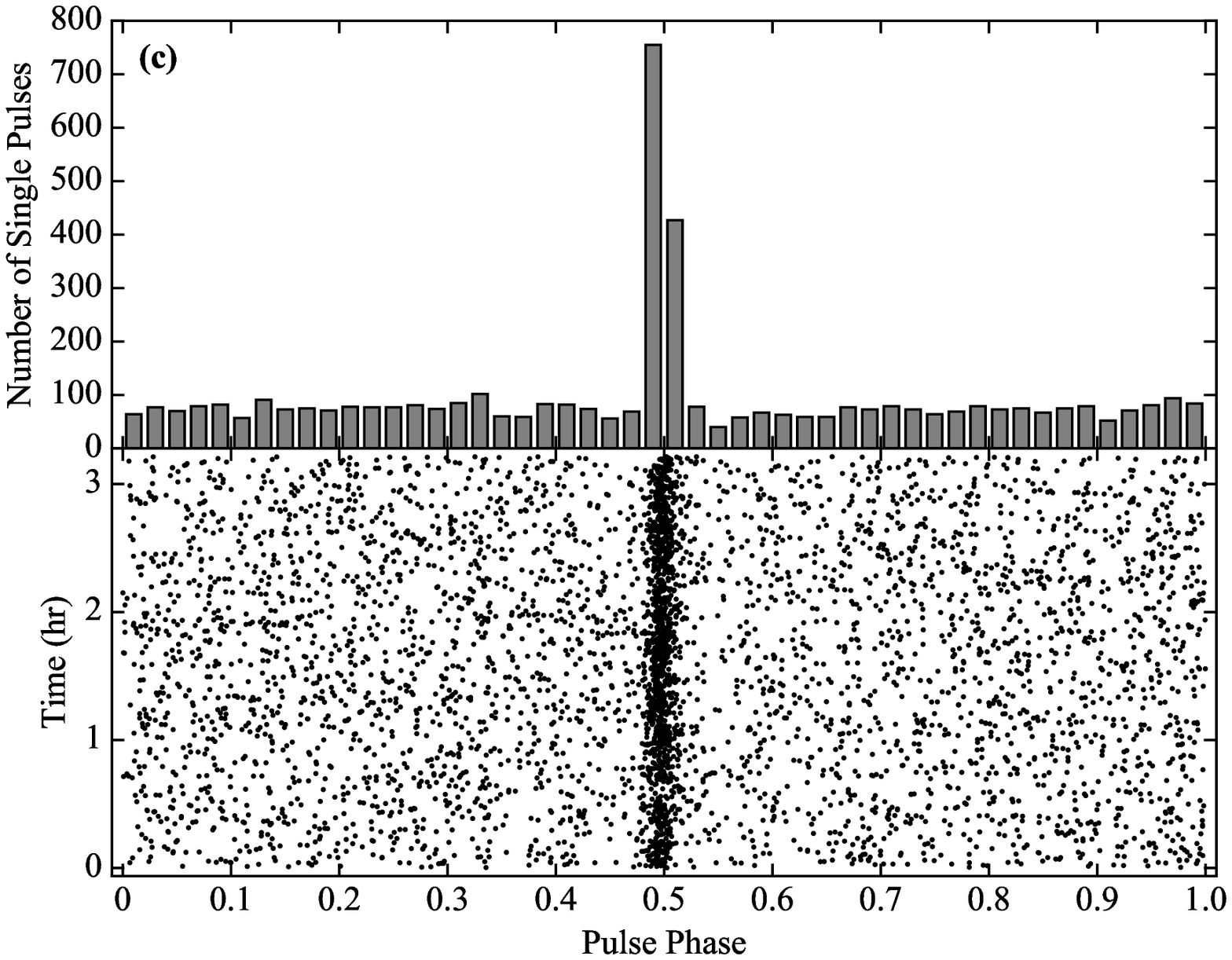}
			\label{Figure:Figure2c}
		}
		
		& &
		
		\subfigure
		{
			\includegraphics[trim=0cm 0cm 0cm 0cm, clip=false, scale=0.4, angle=0]{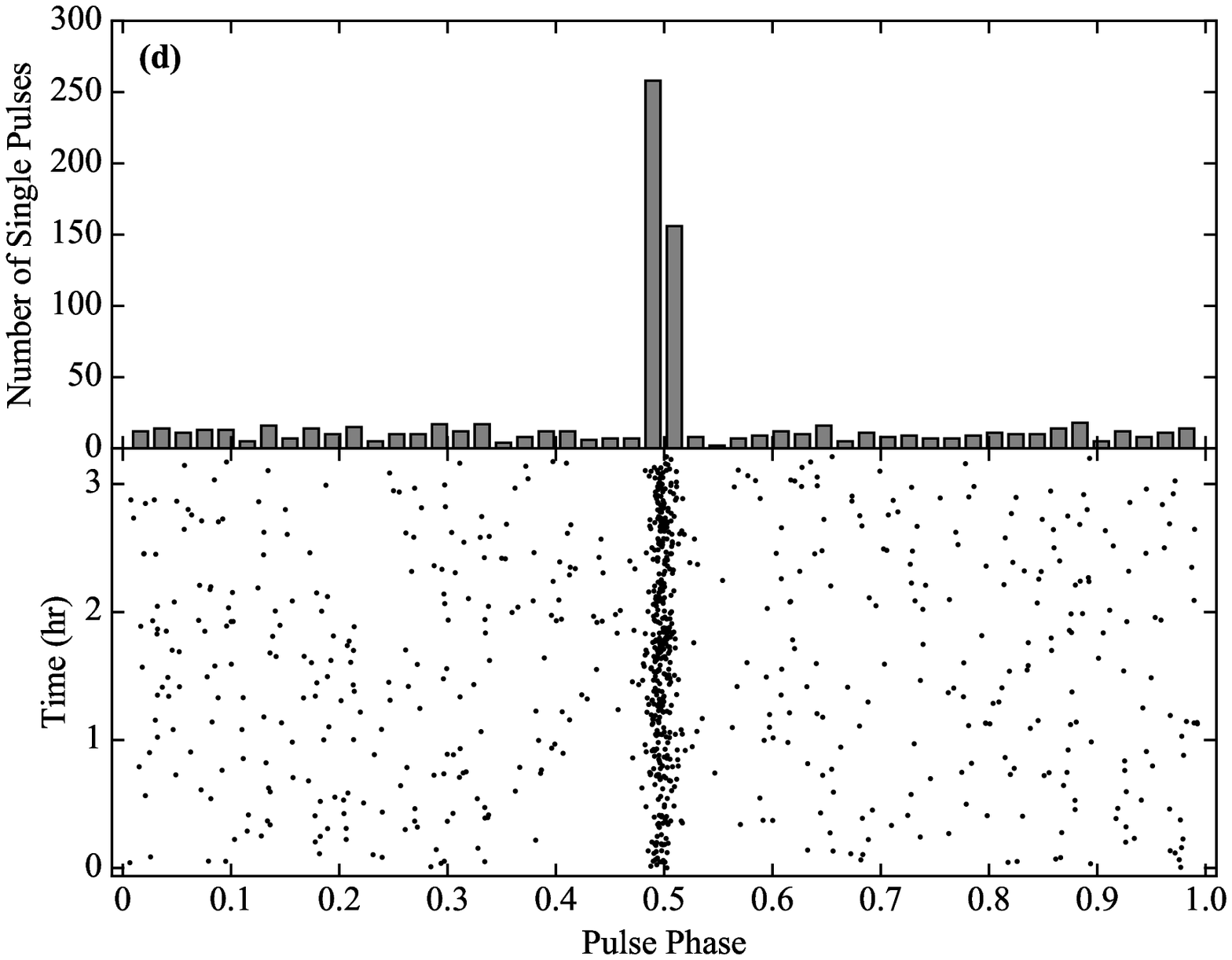}
			\label{Figure:Figure2d}
		}
		
	\end{tabular}
	\caption{Distribution of S-band single pulses in pulse phase during epoch~3 (top row) and epoch~4 (bottom row). The top panels of each figure show the number of single pulses detected in each region of the pulse profile and the bottom panels show the population of single pulses throughout the observation. We show S-band single pulses with SNRs above~4.0 in (a)~and~(c) and single pulses with SNRs above~4.5 are shown in (b)~and~(d).}
	\label{Figure:Figure2}
\end{figure*}

We find a statistically significant population of S-band single pulses from epochs~3 and~4 near the main emission peaks of their pulse profiles. In Figure~\ref{Figure:Figure2}, we show the distribution of single pulses in pulse phase using the measured pulse periods from each epoch. With an SNR threshold of 4.0 in Figure~\ref{Figure:Figure2a} and Figure~\ref{Figure:Figure2c}, which is equivalent to a peak flux density of 0.46\,Jy, we detect 573\,$\pm$\,127 and 1040\,$\pm$\,73~events above the background during epochs~3 and~4, respectively. This corresponds to a single pulse event rate of 1.2\% and 3.7\% per stellar rotation during epochs~3 and~4, respectively. The uncertainties were determined from fitting the background rate independently of pulse phase. Figure~\ref{Figure:Figure2b} and Figure~\ref{Figure:Figure2d} show single pulses with SNRs above 4.5 and peak flux densities greater than~0.51\,Jy. Using this selection criteria, we find 148\,$\pm$\,20 and 393\,$\pm$\,11 single pulses above the background during epochs~3 and~4, respectively, corresponding to a single pulse event rate of 0.3\% and 1.4\% per stellar rotation during the two epochs.  From these event rates, and depending on the chosen value for the SNR threshold cut, we estimate a factor of 3--4 increase in the overall single pulse emission rate in epoch~4 compared to epoch~3. Time of arrivals (ToAs), pulse widths, SNRs, and false alarm probabilities (FAPs) for individual, bright single pulses in these epochs and additional epochs following these observations will be presented in a later paper~\citep{Pearlman2017}.

%%%%%%%%%%%%%%%%%%%%%%%%%%%%%%%%%%%%%%%%%%%%%%%%%%%%%%%%%%%%%%

% Figure 2 - Distribution of Single Pulses

%%%%%%%%%%%%%%%%%%%%%%%%%%%%%%%%%%%%%%%%%%%%%%%%%%%%%%%%%%%%%%

\section{Discussion}
\label{Section:Discussion}

PSR~J1119-6127 is clearly a transition object, i.e. a high-magnetic field neutron star that is normally a rotation-powered pulsar in radio and X-rays, but also shows transient magnetar-like behavior. Such behavior is unlikely to be powered solely by rotation, but also by the release of stored magnetic energy~\citep{Archibald2016a}. This was previously suggested because of its unusual pulsed X-ray emission~\citep{Gonzalez2005}, which was hard to reconcile with the thermal emission from the rotation-powered pulsar. This is now dramatically confirmed by clear magnetar-like outbursts~\citep{Archibald2016a, Gogus2016}. PSR~J1119-6127 now joins PSR~J1846-0258 as a high-magnetic field pulsar with transient magnetar-like behavior~\citep{Kaspi2010, Ng2011}. PSR~J1119-6127 is similar to PSR~J1846-0258 in terms of its field strength and young characteristic age. However, while PSR~J1846-0258 is radio quiet, PSR~J1119-6127 shows radio emission both in its ``quiescent'' rotation-powered state, as well as in its magnetar-like state.

Of particular interest is the observation of multi-peaked S-band radio emission shortly after
the outburst. In its normal rotation-powered state, PSR~J1119-6127 has a single-peaked pulse profile at 1.4\,GHz~\citep{Camillo2000} that is aligned with a single peaked~\citep{Gonzalez2005} broad profile in the 0.5--2\,keV emission band, which is consistent with thermal emission from the polar cap. In $\gamma$-rays, PSR~J1119-6127 also shows a single-peaked profile consistent with outer gap emission~\citep{Parent2011}. Our observation of multi-component emission at S-band shortly after the outburst is indicative of a more complex emission geometry and possibly non-dipolar field components near the neutron star surface. Multi-component emission from PSR~J1119-6127 was seen only once before and immediately after one of PSR~J1119-6127's strong glitches~\citep{Weltevrede2011}, observed only once in 12~years of monitoring.  \citet{Weltevrede2011} concluded that this emission behavior was extremely rare since it was only observed in 0.1\% of their inspected data. In contrast, our observations during epochs~3 and~4 show multiple-peaked emission lasting more than a week. Remarkably, observations in epoch~4 show dramatic changes in both S-band and X-band. The S-band profile seems to be returning to a single peak with reduced emission in the first two emission regions. The X-band emission in epoch~4 is brighter than the previous X-band detection in epoch~3 and seems to be singly peaked.

\citet{Weltevrede2011} also reported the detection of a handful of individual pulses during the glitch recovery phase in 2011. We also observe multiple individual pulses in both observations after the reactivation of pulsed radio emission. Additional high frequency radio observations following epoch~4, which are not presented in this letter, show that the pulse profile at S-band is still evolving. We are continuing to observe PSR~J1119-6127 and are finding that the pulse profile is slowly returning to a single-peaked emission structure. These results, along with a detailed study of the mode changes, will be reported in a later work~\citep{Pearlman2017}.

Four of the 26 known magnetars~\citep{Olausen2014} have been detected in the radio band: XTE~J1810-197, 1E~1547.0-5408, PSR~J1622-4950, and SGR~J1745-2900 \citep{Camilo2006, Camilo2007a, Levin2012, Shannon2013}, and they all show variability in their pulse profiles, radio flux densities, and spectral indices.
Furthermore, both XTE~J1810-197 and 1E~1547.0-5408 have exhibited sudden appearance and then fading of their radio emission over time scales of days to months \citep{Camilo2007b, Camilo2008}. In addition to an evolving pulse profile, we also report a variable emission flux at X-band showing an increase of roughly a factor of 2 during epoch~4 compared to the observed flux during epoch~3.  A similar factor of 3--4 increase is also seen in the single pulse event rate at S-band between these two epochs. In this respect, PSR~J1119-6127 shares many of the properties of the known radio magnetars.  However, while the spectral indices of radio magnetars tend to be quite flat, the spectral index of PSR~J1119-6127 is more similar to the majority of rotation-powered pulsars, perhaps indicative of the transitional nature of this object. 

One possible explanation for the transient radio emission seen in magnetars and similar behavior observed from high magnetic field pulsars, such as PSR~J1119-6127, is the dependence of emission on the conditions of the magnetosphere \citep{Morozova2012,Lin2015}.   In this model,  toroidal oscillations in the star are excited during an outburst, which then modify the magnetospheric structure and allow radio emission to be produced.  \cite{Lin2015} suggested that, after a glitch, stellar oscillations enlarge the polar cap angle and hence also the size of the radiation cone, resulting in multi-component pulse profiles seen along the line of sight to the pulsar.   
The proposed explanation for such variability is intriguing since the recent outburst of PSR~J1119-6127 and some outbursts from soft gamma repeaters (SGRs) and Anomalous X-ray Pulsars (AXPs) have been associated with glitches \citep{Kaspi2003,Dib2014}.
More detailed simulations and long-term monitoring of the pulsar are needed to quantitatively investigate this model.

%%%%%%%%%%%%%%%%%%%%%%%%%%%%%%%%%%%%%%%%%%%%%%%%%%%%%%%%%%%%%%

\section{Conclusion}
\label{Section:Conclusion}

We have carried out radio observations of PSR~J1119-6127 following its recent X-ray outburst. While initial observations failed to detect the presence of pulsed emission, subsequent observations two weeks later show bright detections of the pulsar at S-band and a significant detection at X-band as the S-band pulse profile returns to a single-peaked shape. From these measurements, we were able to estimate a spectral index over a relatively wide range of radio wavelengths. We also detected an unusual multiple-peaked radio profile and single pulse events at S-band. Since this emission behavior is clearly transitory, further radio monitoring of the source is needed to study both the long-term evolution of the pulse profile and the erratic single pulse emission.

%%%%%%%%%%%%%%%%%%%%%%%%%%%%%%%%%%%%%%%%%%%%%%%%%%%%%%%%%%%%%%

\section*{Acknowledgments}

We thank V.~Kaspi and R.~Archibald for alerting us to the initial outburst and for providing an ephemeris for PSR~J1119-6127. We also thank V.~Kaspi for a careful reading of the manuscript and detailed comments. We acknowledge support from the DSN team for scheduling the observations.
A.~B.~Pearlman acknowledges support by the Department of Defense (DoD) through the National Defense Science and Engineering Graduate Fellowship (NDSEG) Program and by the National Science Foundation Graduate Research Fellowship under Grant No.~DGE-1144469. J.~Lippuner acknowledges support from the Jet Propulsion Laboratory Graduate Fellowship program. A portion of this research was performed at the Jet Propulsion Laboratory, California Institute of Technology under a Research and Technology Development Grant and under a contract with the National Aeronautics and Space Administration. Copyright 2016 California Institute of Technology. Government sponsorship acknowledged.

%\bibliography{bib}

\clearpage

\end{document}